\documentclass[a4paper,aps,onecolumn,nofootinbib]{revtex4}
\RequirePackage[colorlinks,hyperindex]{hyperref}
\RequirePackage[english]{babel}
\RequirePackage[latin1]{inputenc}
\RequirePackage[T1]{fontenc}
\RequirePackage{mathrsfs}
\RequirePackage{amsmath}
\RequirePackage{amssymb}
\RequirePackage{amsbsy}
\RequirePackage{color}
\RequirePackage{bm}
\hypersetup{colorlinks=true,breaklinks=true,urlcolor=blue,linkcolor=red}
\begin{document}
\title{\bf{Polar Form of the Dirac Operator in Low-Dimensional Space-Times}}
\author{Luca Fabbri$^{c}$\!\!\! $^{\hbar}$\!\!\! $^{G}$\footnote{luca.fabbri@unige.it},
Giuseppe De Maria$^{c}$\!\!\! $^{\hbar}$\footnote{giuseppe.demaria@edu.unige.it}}
\affiliation{$^{c}$DIME, Universit\`{a} di Genova, Via all'Opera Pia 15, 16145 Genova, ITALY\\
$^{\hbar}$INFN, Sezione di Genova, Via Dodecaneso 33, 16146 Genova, ITALY\\
$^{G}$GNFM, Istituto Nazionale di Alta Matematica, P.le Aldo Moro 5, 00185 Roma, ITALY}
\date{\today}
\begin{abstract}
We express in polar form the Dirac operator as left-multiplication by an assigned matrix: we use this technique to find its right-resolvent for low-dimensional spaces.
\end{abstract}
\maketitle
\section{Introduction}
The polar decomposition of any complex function is the procedure in terms of which it can be written as the product of a module times a phase. When spinors are taken, their being complex-valued allows such a polar decomposition to be made, although the fact that they have two components, accounting for both helicities, makes these new variables mix with each other, when rotations are performed. It is even worse for relativistic spinors, since the fact that they have four components, accounting for both helicities and both chiralities, makes the new variables mix under a Lorentz transformation. Yet, it is still possible, by exercising due care, to make such a polar decomposition also for relativistic spinors, and at the same time have covariance respected. These results are in the literature since long \cite{jl1,jl2}, and while they are not really well-known among physicists, a presentation suitable for physical applications can be found in \cite{Fabbri:2018crr}.

Recently, such a polar decomposition has been lifted to the differential structures \cite{Fabbri:2018crr, Fabbri:2024avj}, in which the action of the covariant derivative is correspondently re-formulated as the left-multiplication by an assigned matrix. Such a fact is particularly important because the mathematical treatment of objects like the Dirac operator, notoriously complicated in general instances, can be brought down to the treatment of a matrix operator, considerably simpler instead.

In this paper, we focus on finding the right-resolvent of the Dirac operator. The resulting expression is more general than the usual propagator of quantum field theory, as we are going to show with an application.

To be more pedagogical, we will perform the analysis in several dimensions.
\section{Euclidean Surfaces}\label{ESu}
We begin from the simplest possible case, that is the Euclidean surface, or equivalently the $(0\!+\!2)$-dimensional case.

In this case the Clifford matrices $\boldsymbol{\gamma}_{a}$ ($\boldsymbol{\gamma}$ standing for Clifford) are defined to verify $\left\{\boldsymbol{\gamma}_{a},\boldsymbol{\gamma}_{b}\right\}\!=\!2\delta_{ab}\mathbb{I}$ where $\delta_{ab}$ is the $2$-dimensional Kronecker delta. The $\boldsymbol{\sigma}_{ab}$ ($\boldsymbol{\sigma}$ standing for spin) are defined as $\boldsymbol{\sigma}_{ab}\!=\!\left[\boldsymbol{\gamma}_{a},\boldsymbol{\gamma}_{b}\right]/4$ as generator of rotations taking place on the surface. The $\boldsymbol{\pi}$ ($\boldsymbol{\pi}$ standing for parity) is defined by $2i\boldsymbol{\sigma}_{ab}\!=\!\varepsilon_{ab}\boldsymbol{\pi}$ in which $\varepsilon_{ab}$ is the Levi-Civita completely antisymmetric pseudo-tensor. Moreover, we have the relation $\boldsymbol{\gamma}^{a}\boldsymbol{\pi}\!=\!i\varepsilon^{ab}\boldsymbol{\gamma}_{b}$ specifically in this dimension.

For Dirac spinors in this dimension, the only possible rotation can be easily computed, and with it, it is not difficult to see that the most general spinor can always be written in the so-called polar decomposition according to
\begin{eqnarray}
&\!\psi\!=\!\phi\ e^{-\frac{1}{2}\eta\boldsymbol{\pi}}
\boldsymbol{L}^{-1}\left(\!\begin{tabular}{c}
$1$\\
$1$
\end{tabular}\!\right)
\label{spinor2}
\end{eqnarray}
for some $\phi$ and $\eta$ real and with $\boldsymbol{L}^{-1}$ being the complex representation of the rotation possibly containing also a global unitary phase. This form can be obtained only up to the $\psi\!\rightarrow\!\boldsymbol{\pi}\psi$ discrete transformation. For a demonstration of all these statements, see Appendix \ref{app1}. Then, with $\overline{\psi}\!=\!\psi^{\dagger}$, we have
\begin{eqnarray}
&\overline{\psi}\psi\!=\!\phi^{2}\cosh{\eta}\ \ \ \ \ \ \ \
\overline{\psi}\boldsymbol{\pi}\psi\!=\!-\phi^{2}\sinh{\eta}\\
&\overline{\psi}\boldsymbol{\gamma}^{a}\psi\!=\!\phi^{2}s^{a}
\end{eqnarray}
so that $\phi$ and $\eta$ are scalars and $s^{a}$ is the spin vector. It verifies $s_{a}s^{a}\!=\!1$ as normalization.

The covariant derivative of spinors in polar form can always be written as
\begin{eqnarray}
&\boldsymbol{\nabla}_{\mu}\psi\!=\!(\nabla_{\mu}\ln{\phi}\mathbb{I}
\!-\!\frac{1}{2}\nabla_{\mu}\eta\boldsymbol{\pi}
\!-\!iP_{\mu}\mathbb{I}\!-\!\frac{1}{2}R_{ab\mu}\boldsymbol{\sigma}^{ab})\psi
\label{decspinder2}
\end{eqnarray}
for some gauge-invariant vector $P_{\mu}$ called momentum and for some real tensor $R_{ij\mu}$ called tensorial connection. For a proof, see Appendix \ref{app2}. Then
\begin{eqnarray}
&\nabla_{\mu}s_{i}\!=\!s^{j}R_{ji\mu}
\end{eqnarray}
is a general identity.

By introducing the vector-valued spinorial matrix
\begin{eqnarray}
&\boldsymbol{M}_{\mu}\!=\!\nabla_{\mu}\ln{\phi}\mathbb{I}
\!-\!\frac{1}{2}\nabla_{\mu}\eta\boldsymbol{\pi}
\!-\!iP_{\mu}\mathbb{I}\!-\!\frac{1}{2}R_{ab\mu}\boldsymbol{\sigma}^{ab}
\label{Mi2}
\end{eqnarray}
we can write
\begin{gather}
\boldsymbol{\nabla}_{\mu}\psi\!=\!\boldsymbol{M}_{\mu}\psi
\label{spinder2}
\end{gather}
showing that the covariant derivative acts as left-multiplication by the associated $\boldsymbol{M}_{\mu}$ matrix.

This turns out to be particularly useful in the computation of the right-resolvent operator. The right-resolvent of the Dirac operator is the right-inverse of the Dirac equation, that is the operator $\boldsymbol{R}$ verifying $(i\boldsymbol{\gamma}^{\mu}\boldsymbol{\nabla}_{\mu}\!-\!m)\boldsymbol{R}\!=\!\mathbb{I}$ in the operatorial sense: equivalently $(i\boldsymbol{\gamma}^{\mu}\boldsymbol{\nabla}_{\mu}\!-\!m)(\boldsymbol{R}\psi)\!=\!\psi$ for any spinor field $\psi$ in general. Explicitly, this implies that
\begin{gather}
i\boldsymbol{\gamma}^{\mu}\nabla_{\mu}\boldsymbol{R}
\!+\!i\nabla_{\mu}\ln{\phi}\boldsymbol{\gamma}^{\mu}\boldsymbol{R}
\!-\!\frac{i}{2}\nabla_{\mu}\eta\boldsymbol{\gamma}^{\mu}\boldsymbol{R}\boldsymbol{\pi}
\!+\!P_{\mu}\boldsymbol{\gamma}^{\mu}\boldsymbol{R}
\!-\!\frac{i}{2}R_{ab\mu}\boldsymbol{\gamma}^{\mu}\boldsymbol{R}\boldsymbol{\sigma}^{ab}
\!-\!m\boldsymbol{R}\!-\!\mathbb{I}\!=\!0
\label{resolv2}
\end{gather}
in which (\ref{Mi2}) was used. We look for a solution for $\boldsymbol{R}$, which, in general, can be taken in the form
\begin{gather}
\boldsymbol{R}\!=\!(S\mathbb{I}
\!+\!V_{a}\boldsymbol{\gamma}^{a}
\!+\!P\boldsymbol{\pi})e^{\frac{1}{2}\eta\boldsymbol{\pi}}\phi^{-1}\label{R2}
\end{gather}
where $S$, $V_{a}$ and $P$ have to be found by integrability conditions (the scaling factor $e^{\frac{1}{2}\eta\boldsymbol{\pi}}\phi^{-1}$ does not account for loss of generality). Plugging into (\ref{resolv2}) the expression (\ref{R2}) we find
\begin{eqnarray}
\nonumber
&i\boldsymbol{\gamma}^{\mu}(\nabla_{\mu}S\mathbb{I}
\!+\!\nabla_{\mu}V_{a}\boldsymbol{\gamma}^{a}
\!+\!\nabla_{\mu}P\boldsymbol{\pi})
\!+\!P_{\mu}\boldsymbol{\gamma}^{\mu}(S\mathbb{I}
\!+\!V_{a}\boldsymbol{\gamma}^{a}
\!+\!P\boldsymbol{\pi})-\\
&-\frac{i}{2}R_{ij\mu}\boldsymbol{\gamma}^{\mu}(S\mathbb{I}
\!+\!V_{a}\boldsymbol{\gamma}^{a}
\!+\!P\boldsymbol{\pi})\boldsymbol{\sigma}^{ij}
\!-\!m(S\mathbb{I}
\!+\!V_{a}\boldsymbol{\gamma}^{a}
\!+\!P\boldsymbol{\pi})\!-\!\phi e^{-\frac{1}{2}\eta\boldsymbol{\pi}}\!=\!0:
\end{eqnarray}
performing the products employing the properties of the Clifford matrices, one straightforwardly gets
\begin{eqnarray}
\nonumber
&[i\nabla_{a}V^{a}\!+\!P_{a}V^{a}
\!+\!\frac{i}{4}\varepsilon^{ij}R_{ijk}V_{a}\varepsilon^{ka}
\!-\!mS\!-\!\phi\cosh{(\eta/2)}]\mathbb{I}+\\
\nonumber
&+[i\nabla_{a}S\!+\!P_{a}S
\!-\!\frac{i}{4}\varepsilon_{ij}R^{ijk}S\varepsilon_{ka}
\!-\!\varepsilon_{ka}\nabla^{k}P\!+\!i\varepsilon_{ka}P^{k}P
\!-\!\frac{1}{4}\varepsilon^{ij}R_{ija}P
\!-\!mV_{a}]\boldsymbol{\gamma}^{a}-\\
&-i[i\varepsilon^{ka}\nabla_{k}V_{a}\!+\!\varepsilon^{ka}P_{k}V_{a}
-\frac{i}{4}\varepsilon^{ij}R_{ijk}V^{k}
\!-\!imP\!+\!i\phi\sinh{(\eta/2)}]\boldsymbol{\pi}\!=\!0
\end{eqnarray}
and because all matrices are linearly independent, it follows that
\begin{eqnarray}
&i\nabla_{a}V^{a}\!+\!P_{a}V^{a}
\!+\!\frac{i}{4}\varepsilon^{ij}R_{ijk}V_{a}\varepsilon^{ka}
\!-\!mS\!-\!\phi\cosh{(\eta/2)}\!=\!0
\end{eqnarray}
\begin{eqnarray}
&i\varepsilon^{ka}\nabla_{k}V_{a}\!+\!\varepsilon^{ka}P_{k}V_{a}
-\frac{i}{4}\varepsilon^{ij}R_{ijk}V^{k}
\!-\!imP\!+\!i\phi\sinh{(\eta/2)}\!=\!0
\end{eqnarray}
\begin{eqnarray}
&i\nabla_{a}S\!+\!P_{a}S
\!-\!\frac{i}{4}\varepsilon_{ij}R^{ijk}S\varepsilon_{ka}
\!-\!\varepsilon_{ka}\nabla^{k}P\!+\!i\varepsilon_{ka}P^{k}P
\!-\!\frac{1}{4}\varepsilon^{ij}R_{ija}P
\!-\!mV_{a}\!=\!0
\end{eqnarray}
as the integrability conditions that must be satisfied to obtain the right-resolvent on Euclidean surfaces.
\section{Lorentzian Surfaces}\label{LSu}
Our next example is just the case above plus one temporal dimension, described by the $(1\!+\!2)$-dimensional geometry.

In this case the Clifford matrices $\boldsymbol{\gamma}_{a}$ are defined as above as $\left\{\boldsymbol{\gamma}_{a},\boldsymbol{\gamma}_{b}\right\}\!=\!2\eta_{ab}\mathbb{I}$ but now $\eta_{ab}$ is the $(1\!+\!2)$-dimensional Minkowski matrix. The $\boldsymbol{\sigma}_{ab}$ are defined as above. The $\boldsymbol{\pi}$ is not defined. We have that $2\boldsymbol{\sigma}^{ab}\!=\!i\varepsilon^{abc}\boldsymbol{\gamma}_{c}$ in which $\varepsilon_{abc}$ is the Levi-Civita completely antisymmetric pseudo-tensor, showing that the $\boldsymbol{\sigma}^{ab}$ are replaceable with Clifford matrices.

For Dirac spinors the polar decomposition is given by
\begin{eqnarray}
&\psi\!=\!\phi \boldsymbol{L}^{-1}\left(\!\begin{tabular}{c}
$0$\\
$1$
\end{tabular}\!\right)\ \ \ \ \ \ \ \ \mathrm{or}\ \ \ \ \ \ \ \
\psi\!=\!\phi \boldsymbol{L}^{-1}\left(\!\begin{tabular}{c}
$1$\\
$0$
\end{tabular}\!\right)
\label{spinor3}
\end{eqnarray}
with $\phi$ a real function, and for some $\boldsymbol{L}$ with the structure of a complex Lorentz transformation possibly complemented by a global unitary phase. For a proof of these statements, see Appendix \ref{app1}. In polar form, with $\overline{\psi}\!=\!\psi^{\dagger}\boldsymbol{\gamma}_{0}$, we have
\begin{eqnarray}
&\overline{\psi}\psi\!=\!\phi^{2}\\
&\overline{\psi}\boldsymbol{\gamma}^{a}\psi\!=\!\phi^{2}u^{a}
\end{eqnarray}
showing that $\phi$ is a scalar and $u^{a}$ is the velocity vector. It verifies $u_{a}u^{a}\!=\!1$ as normalization.

The covariant derivative of spinors in polar form can be expressed according to
\begin{eqnarray}
&\boldsymbol{\nabla}_{\mu}\psi\!=\!(\nabla_{\mu}\ln{\phi}\mathbb{I}
\!-\!iP_{\mu}\mathbb{I}\!-\!\frac{1}{2}R_{ab\mu}\boldsymbol{\sigma}^{ab})\psi
\label{decspinder3}
\end{eqnarray}
with $P_{\mu}$ momentum and $R_{ij\mu}$ tensorial connection \ref{app2}. Then
\begin{eqnarray}
&\nabla_{\mu}u_{i}\!=\!u^{j}R_{ji\mu}
\end{eqnarray}
is a general identity.

By introducing the vector-valued spinorial matrix
\begin{eqnarray}
&\boldsymbol{M}_{\mu}\!=\!\nabla_{\mu}\ln{\phi}\mathbb{I}
\!-\!iP_{\mu}\mathbb{I}\!-\!\frac{1}{2}R_{ab\mu}\boldsymbol{\sigma}^{ab}
\label{Mi3}
\end{eqnarray}
we can write
\begin{gather}
\boldsymbol{\nabla}_{\mu}\psi\!=\!\boldsymbol{M}_{\mu}\psi
\label{spinder3}
\end{gather}
showing that the covariant derivative acts as left-multiplication by the associated $\boldsymbol{M}_{\mu}$ matrix.

In this case, the right-resolvent of the Dirac operator is expressed by the condition
\begin{gather}
i\boldsymbol{\gamma}^{\mu}\nabla_{\mu}\boldsymbol{R}
\!+\!i\nabla_{\mu}\ln{\phi}\boldsymbol{\gamma}^{\mu}\boldsymbol{R}
\!+\!P_{\mu}\boldsymbol{\gamma}^{\mu}\boldsymbol{R}
\!-\!\frac{i}{2}R_{ab\mu}\boldsymbol{\gamma}^{\mu}\boldsymbol{R}\boldsymbol{\sigma}^{ab}
\!-\!m\boldsymbol{R}\!-\!\mathbb{I}\!=\!0
\label{resolv3}
\end{gather}
in which (\ref{Mi3}) was used. We look for a solution for $\boldsymbol{R}$, which, in general, can be taken in the form
\begin{gather}
\boldsymbol{R}\!=\!(S\mathbb{I}
\!+\!V_{a}\boldsymbol{\gamma}^{a})\phi^{-1}\label{R3}
\end{gather}
where $S$ and $V_{a}$ have to be found by integrability conditions (the scaling $\phi^{-1}$ accounts for no loss of generality). Upon substitution in (\ref{resolv3}) of (\ref{R3}) we find
\begin{gather}
i\boldsymbol{\gamma}^{\mu}(\nabla_{\mu}S\mathbb{I}
\!+\!\nabla_{\mu}V_{a}\boldsymbol{\gamma}^{a})
\!+\!P_{\mu}\boldsymbol{\gamma}^{\mu}(S\mathbb{I}
\!+\!V_{a}\boldsymbol{\gamma}^{a})
\!-\!\frac{i}{2}R_{ij\mu}\boldsymbol{\gamma}^{\mu}(S\mathbb{I}
\!+\!V_{a}\boldsymbol{\gamma}^{a})\boldsymbol{\sigma}^{ij}
\!-\!m(S\mathbb{I}
\!+\!V_{a}\boldsymbol{\gamma}^{a})\!-\!\phi\mathbb{I}\!=\!0:
\end{gather}
as before, properties of the Clifford matrices like $\boldsymbol{\gamma}^{a}\boldsymbol{\gamma}^{b}\boldsymbol{\gamma}^{c}\!=\!\boldsymbol{\gamma}^{a}\eta^{bc}\!-\!\boldsymbol{\gamma}^{b}\eta^{ac}\!+\!\boldsymbol{\gamma}^{c}\eta^{ab}\!+\!i\varepsilon^{abc}\mathbb{I}$ lead to
\begin{eqnarray}
\nonumber
&[i\nabla_{k}V^{k}\!+\!P_{k}V^{k}\!-\!\frac{i}{2}R_{k}V^{k}
\!+\!\frac{1}{4}\varepsilon^{ijk}R_{ijk}S\!-\!mS\!-\!\phi]\mathbb{I}+\\
&+[i\nabla_{a}S\!+\!P_{a}S\!+\!\frac{i}{2}R_{a}S
\!-\!\nabla^{i}V^{j}\varepsilon_{ija}\!+\!iP^{i}V^{j}\varepsilon_{ija}
\!+\!\frac{1}{4}\varepsilon^{ijk}R_{ija}V_{k}\!-\!\frac{1}{4}\varepsilon^{ijk}R_{ijk}V_{a}
\!+\!\frac{1}{4}\varepsilon_{ija}R^{ijk}V_{k}\!-\!mV_{a}]\boldsymbol{\gamma}^{a}\!=\!0
\end{eqnarray}
and because of the linear independence of the Clifford matrices we can write
\begin{eqnarray}
&i\nabla_{k}V^{k}\!+\!P_{k}V^{k}\!-\!\frac{i}{2}R_{k}V^{k}
\!+\!\frac{1}{4}\varepsilon^{ijk}R_{ijk}S\!-\!mS\!-\!\phi\!=\!0
\end{eqnarray}
\begin{eqnarray}
&i\nabla_{a}S\!+\!P_{a}S\!+\!\frac{i}{2}R_{a}S
\!-\!\nabla^{i}V^{j}\varepsilon_{ija}\!+\!iP^{i}V^{j}\varepsilon_{ija}
\!+\!\frac{1}{4}\varepsilon^{ijk}R_{ija}V_{k}\!-\!\frac{1}{4}\varepsilon^{ijk}R_{ijk}V_{a}
\!+\!\frac{1}{4}\varepsilon_{ija}R^{ijk}V_{k}\!-\!mV_{a}\!=\!0
\end{eqnarray}
as the integrability conditions on Lorentzian surfaces.
\section{Lorentzian Spaces}\label{LSp}
Our final case is the physical space-time, described by the $(1\!+\!3)$-dimensional geometry.

In this case the Clifford matrices $\boldsymbol{\gamma}_{a}$ are defined as above but now with $\eta_{ab}$ as the $(1\!+\!3)$-dimensional Minkowskian matrix. The $\boldsymbol{\sigma}_{ab}$ are defined as above. The $\boldsymbol{\pi}$ is implicitly defined by $2i\boldsymbol{\sigma}_{ab}\!=\!\varepsilon_{abcd}\boldsymbol{\pi}\boldsymbol{\sigma}^{cd}$ where $\varepsilon_{abcd}$ is the Levi-Civita completely antisymmetric pseudo-tensor (from this expression, after multiplying by $\boldsymbol{\sigma}_{ij}\varepsilon^{abij}$ and performing relatively straightforward computations, it is possible to see that in particular $\boldsymbol{\gamma}_{a}\boldsymbol{\gamma}_{b}\boldsymbol{\gamma}_{i}\boldsymbol{\gamma}_{j}\varepsilon^{abij}\!=\!24i\boldsymbol{\pi}$, which is still in a covariant form, but one for which the matrix $\boldsymbol{\pi}$ has been expressed explicitly).

For the case of physical Dirac spinors the polar decomposition is as
\begin{eqnarray}
&\!\psi\!=\!\phi e^{-\frac{i}{2}\beta\boldsymbol{\pi}}
\boldsymbol{L}^{-1}\left(\!\begin{tabular}{c}
$1$\\
$0$\\
$1$\\
$0$
\end{tabular}\!\right)\ \ \ \ \ \ \ \ \mathrm{or}\ \ \ \ \ \ \ \
\psi\!=\!\phi \boldsymbol{L}^{-1}\left(\!\begin{tabular}{c}
$0$\\
$1$\\
$0$\\
$1$
\end{tabular}\!\right)
\label{spinor4}
\end{eqnarray}
in terms of the scalar $\phi$ called module and the pseudo-scalar $\beta$ called chiral angle, and for some $\boldsymbol{L}$ with the structure of a spinor transformation. These two forms are defined up to the $\psi\!\rightarrow\!\boldsymbol{\pi}\psi$ discrete transformation. For a demonstration of all these statements, see Appendix \ref{app1}. In polar form, and with $\overline{\psi}\!=\!\psi^{\dagger}\boldsymbol{\gamma}_{0}$, one has
\begin{gather}
\overline{\psi}\psi\!=\!2\phi^{2}\cos{\beta}\ \ \ \ \ \ \ \ \ \ \ \ \ \ \ \
i\overline{\psi}\boldsymbol{\pi}\psi\!=\!2\phi^{2}\sin{\beta}\\
\overline{\psi}\boldsymbol{\gamma}^{i}\psi\!=\!2\phi^{2}u^{i}\ \ \ \ \ \ \ \ \ \ \ \ \ \ \ \
\overline{\psi}\boldsymbol{\gamma}^{i}\boldsymbol{\pi}\psi\!=\!2\phi^{2}s^{i}
\end{gather}
with the velocity vector and spin axial-vector. They verify $u^{a}s_{a}\!=\!0$ and $u^{a}u_{a}\!=\!-s_{a}s^{a}\!=\!1$ as ortho-normalization.

When the covariant derivative is computed for spinor fields in polar form we have
\begin{eqnarray}
&\boldsymbol{\nabla}_{\mu}\psi\!=\!(\nabla_{\mu}\ln{\phi}\mathbb{I}
\!-\!\frac{i}{2}\nabla_{\mu}\beta\boldsymbol{\pi}
\!-\!iP_{\mu}\mathbb{I}\!-\!\frac{1}{2}R_{ab\mu}\boldsymbol{\sigma}^{ab})\psi
\label{decspinder4}
\end{eqnarray}
with $P_{\mu}$ the momentum and $R_{ij\mu}$ the tensorial connection \ref{app2}. Then
\begin{gather}
\nabla_{\mu}s_{i}\!=\!s^{j}R_{ji\mu}\ \ \ \ \ \ \ \ \mathrm{and}
\ \ \ \ \ \ \ \ \nabla_{\mu}u_{i}\!=\!u^{j}R_{ji\mu}
\end{gather}
are general identities.

By introducing the vector-valued spinorial matrix
\begin{eqnarray}
&\boldsymbol{M}_{\mu}\!=\!\nabla_{\mu}\ln{\phi}\mathbb{I}
\!-\!\frac{i}{2}\nabla_{\mu}\beta\boldsymbol{\pi}
\!-\!iP_{\mu}\mathbb{I}\!-\!\frac{1}{2}R_{ab\mu}\boldsymbol{\sigma}^{ab}
\label{Mi4}
\end{eqnarray}
we can write
\begin{gather}
\boldsymbol{\nabla}_{\mu}\psi\!=\!\boldsymbol{M}_{\mu}\psi
\label{spinder4}
\end{gather}
showing that the covariant derivative acts as left-multiplication by the associated $\boldsymbol{M}_{\mu}$ matrix.

The right-resolvent of the Dirac operator is the operator $\boldsymbol{R}$ verifying the condition
\begin{eqnarray}
i\boldsymbol{\gamma}^{\mu}\nabla_{\mu}\boldsymbol{R}
\!+\!i\nabla_{\mu}\ln{\phi}\boldsymbol{\gamma}^{\mu}\boldsymbol{R}
\!+\!\frac{1}{2}\nabla_{\mu}\beta\boldsymbol{\gamma}^{\mu}\boldsymbol{R}\boldsymbol{\pi}
\!-\!\frac{i}{2}R_{ab\mu}\boldsymbol{\gamma}^{\mu}\boldsymbol{R}\boldsymbol{\sigma}^{ab}
\!+\!P_{\mu}\boldsymbol{\gamma}^{\mu}\boldsymbol{R}
\!-\!m\boldsymbol{R}\!-\!\mathbb{I}\!=\!0\label{resolv4}
\end{eqnarray}
in which (\ref{Mi4}) was used. We look for a solution for $\boldsymbol{R}$, which can be taken in the form
\begin{equation}
\boldsymbol{R}\!=\!(S\mathbb{I}
\!+\!iP\boldsymbol{\pi}
\!+\!V_{a}\boldsymbol{\gamma}^{a}
\!+\!A_{a}\boldsymbol{\gamma}^{a}\boldsymbol{\pi}
\!+\!iT_{ab}\boldsymbol{\sigma}^{ab})e^{\frac{i}{2}\beta\boldsymbol{\pi}}\phi^{-1}\label{R4}
\end{equation}
where $S$, $P$, $V_{a}$, $A_{a}$ and $T_{ab}$ have to be found by integrability conditions (the scaling factor $e^{\frac{i}{2}\beta\boldsymbol{\pi}}\phi^{-1}$ does not account for any loss of generality). Plugging in (\ref{resolv4}) the expression (\ref{R4}) we get
\begin{eqnarray}
\nonumber
&i\boldsymbol{\gamma}^{\mu}(\nabla_{\mu}S\mathbb{I}
\!+\!i\nabla_{\mu}P\boldsymbol{\pi}
\!+\!\nabla_{\mu}V_{a}\boldsymbol{\gamma}^{a}
\!+\!\nabla_{\mu}A_{a}\boldsymbol{\gamma}^{a}\boldsymbol{\pi}
\!+\!i\nabla_{\mu}T_{ab}\boldsymbol{\sigma}^{ab})-\\
\nonumber
&-\frac{i}{2}R_{cd\mu}\boldsymbol{\gamma}^{\mu}(S\mathbb{I}
\!+\!iP\boldsymbol{\pi}
\!+\!V_{a}\boldsymbol{\gamma}^{a}
\!+\!A_{a}\boldsymbol{\gamma}^{a}\boldsymbol{\pi}
\!+\!iT_{ab}\boldsymbol{\sigma}^{ab})\boldsymbol{\sigma}^{cd}+\\
\nonumber
&+P_{\mu}\boldsymbol{\gamma}^{\mu}(S\mathbb{I}
\!+\!iP\boldsymbol{\pi}
\!+\!V_{a}\boldsymbol{\gamma}^{a}
\!+\!A_{a}\boldsymbol{\gamma}^{a}\boldsymbol{\pi}
\!+\!iT_{ab}\boldsymbol{\sigma}^{ab})-\\
&-m(S\mathbb{I}
\!+\!iP\boldsymbol{\pi}
\!+\!V_{a}\boldsymbol{\gamma}^{a}
\!+\!A_{a}\boldsymbol{\gamma}^{a}\boldsymbol{\pi}
\!+\!iT_{ab}\boldsymbol{\sigma}^{ab})
\!-\!\phi e^{-\frac{i}{2}\beta\boldsymbol{\pi}}\!=\!0\label{exp}:
\end{eqnarray}
computations in this case are much more difficult since product of Clifford matrices go up to involving four, or even more, factors, but it is still straightforward to prove that (\ref{exp}) can be expanded into the sum of five terms, proportional to $\mathbb{I}$, $\boldsymbol{\gamma}^{i}$, $\boldsymbol{\sigma}^{ab}$, $\boldsymbol{\gamma}^{i}\boldsymbol{\pi}$, $\boldsymbol{\pi}$, which are known to be linearly independent. Consequently, the five terms must be independently equal to zero, and this leads to the five expressions
\begin{eqnarray}
&i\nabla_{k}V^{k}\!+\!P_{k}V^{k}\!-\!\frac{i}{2}R_{k}V^{k}\!-\!\frac{1}{2}B_{k}A^{k}\!-\!mS
\!-\!\phi\cos{(\beta/2)}\!=\!0\label{int1}
\end{eqnarray}
\begin{eqnarray}
&i\nabla_{k}A^{k}\!+\!P_{k}A^{k}\!-\!\frac{i}{2}R_{k}A^{k}\!-\!\frac{1}{2}B_{k}V^{k}\!-\!imP
\!+\!i\phi\sin{(\beta/2)}\!=\!0\label{int2}
\end{eqnarray}
\begin{eqnarray}
\nonumber
&i\nabla^{a}S\!+\!P^{a}S\!+\!\frac{i}{2}R^{a}S\!+\!\frac{i}{2}B^{a}P-\\
&-\nabla_{k}T^{ka}\!+\!iP_{k}T^{ka}
\!+\!\frac{1}{2}R_{k}T^{ka}\!-\!\frac{1}{4}B_{i}T_{jk}\varepsilon^{ijka}
\!-\!\frac{1}{2}R^{kja}T_{kj}\!-\!mV^{a}\!=\!0\label{int3}
\end{eqnarray}
\begin{eqnarray}
\nonumber
&i\nabla^{a}P\!+\!P^{a}P\!+\!\frac{i}{2}R^{a}P\!-\!\frac{i}{2}B^{a}S+\\
&+\frac{1}{2}\nabla_{i}T_{jk}\varepsilon^{ijka}\!-\!\frac{i}{2}P_{i}T_{jk}\varepsilon^{ijka}
\!-\!\frac{1}{4}R_{i}T_{jk}\varepsilon^{ijka}\!-\!\frac{1}{2}B_{i}T^{ia}
\!+\!\frac{1}{4}R^{\mu\nu a}T^{ij}\varepsilon^{ij\mu\nu}\!+\!imA^{a}\!=\!0\label{int4}
\end{eqnarray}
\begin{eqnarray}
\nonumber
&i\nabla_{[a}V_{b]}\!+\!P_{[a}V_{b]}
\!+\!\frac{i}{2}R_{[a}V_{b]}\!-\!\frac{i}{2}B^{i}V^{j}\varepsilon_{ijab}
\!-\!iR_{abk}V^{k}+\\
&+\nabla^{i}A^{j}\varepsilon_{ijab}\!-\!iP^{i}A^{j}\varepsilon_{ijab}
\!+\!\frac{1}{2}R^{i}A^{j}\varepsilon_{ijab}\!+\!\frac{1}{2}B_{[a}A_{b]}
\!-\!\frac{1}{2}R^{ijk}A_{k}\varepsilon_{ijab}\!-\!imT_{ab}\!=\!0\label{int5}
\end{eqnarray}
as integrability conditions. A shorter approach would be to have (\ref{exp}) multiplied in turn by $\mathbb{I}$, $\boldsymbol{\gamma}^{i}$, $\boldsymbol{\sigma}^{ab}$, $\boldsymbol{\gamma}^{i}\boldsymbol{\pi}$, $\boldsymbol{\pi}$, in each case taking the trace, and the same integrability conditions above would have been obtained (for details, see \ref{app3}).
\section{An Application}
A special solution of the above integrability conditions is obtained in the situation in which $\beta\!=\!0$ and $R_{ij\mu}\!=\!0$ and in which $P\!=\!A^{a}\!=\!0$ with $S$, $V_{a}$, $T_{ab}$ real: in such a circumstance, we have in fact
\begin{eqnarray}
&\nabla_{k}V^{k}\!=\!0\ \ \ \ \ \ \ \ \ \ \ \ \ \ \ \ P_{k}V^{k}\!-\!mS\!-\!\phi\!=\!0\\
&\nabla^{a}S\!+\!P_{k}T^{ka}\!=\!0\ \ \ \ \ \ \ \ \ \ \ \ \ \ \ \
\nabla_{k}T^{ka}\!+\!mV^{a}\!-\!P^{a}S\!=\!0\\
&P_{i}T_{jk}\varepsilon^{ijka}\!=\!0\ \ \ \ \ \ \ \
\ \ \ \ \ \ \ \ \nabla_{i}T_{jk}\varepsilon^{ijka}\!=\!0\\
&\nabla_{[a}V_{b]}\!-\!mT_{ab}\!=\!0
\ \ \ \ \ \ \ \ \ \ \ \ \ \ \ \ P_{[a}V_{b]}\!=\!0
\end{eqnarray}
from which we see that $mT_{ab}\!=\!\nabla_{[a}V_{b]}$ with $V_{a}\!=\!P_{a}\varphi$ for $\varphi$ to be found. As a consequence $\nabla_{i}T_{jk}\varepsilon^{ijka}\!=\!0$ is identically verified and $P_{i}T_{jk}\varepsilon^{ijka}\!=\!0$ has a structure analogous to that of the Pontryagin topological current. This is in fact the case when $V^{i}\!=\!qa^{i}$ (with $a^{i}$ gauge potential) because in this case $qF_{ab}\!=\!mT_{ab}$ (with $F_{ab}$ gauge strength). For such a gauge potential, picking $S\!=\!m\varphi$ gives $\nabla_{k}F^{ki}\!=\!0$ as well as
\begin{align}
&m^{2}\nabla^{i}\varphi^{2}\!=\!2q^{2}F^{ik}a_{k}\label{cond}
\end{align}
with $\nabla_{k}a^{k}\!=\!0$ and $\phi\!=\!\varphi(P_{k}P^{k}\!-\!m^{2})$ as final conditions. The right-resolvent is therefore
\begin{align}
\boldsymbol{R}\!=\!(P_{i}P^{i}\!-\!m^{2})^{-1}\left(m\mathbb{I}\!+\!P_{a}\boldsymbol{\gamma}^{a}
\!+\!i\frac{q}{m}F_{ab}\boldsymbol{\sigma}^{ab}\varphi^{-1}\right)
\label{R4special}
\end{align}
once $\varphi$ is given as solution of (\ref{cond}), with $F_{ik}\!=\!\partial_{[i}a_{k]}$ and $\nabla^{2}a^{i}\!=\!0$ identically. Therefore, $a^{i}$ can indeed be interpreted as the gauge potential of an electrodynamic field verifying, in Lorenz gauge, the Maxwell equations, in vacuum.

With no electrodynamics, $\boldsymbol{R}\!=\!(P^{2}\!-\!m^{2})^{-1}\left(m\mathbb{I}\!+\!P_{a}\boldsymbol{\gamma}^{a}\right)$, which is the free fermion propagator that is normally used in quantum field theory. What this shows is that (\ref{R4}) is indeed more general than the quantum field propagator.

The next question is now under what circumstances the five integrability conditions (\ref{int1}-\ref{int5}) admit solution, and in the affirmative case, what will it be its physical significance. We leave this discussion to a following work.
\section{Conclusion}
In this article, we have considered the covariant derivative of the spinor field in polar form, which turned out to be expressed as the left-multiplication by a given matrix. We have used this fact to find the right-resolvent of the Dirac operator, and we have discussed, through an example, that our result is in fact more general then the one of QFT.

The presentation covered all low-dimensional spaces, starting from $2$- and $3$-dimensions, where computations were straightforward, so that the reader would see the main idea without being overwhelmed by technicalities in calculations.

The presented result is relevant in the context of spectral theory since, to our knowledge, this is the first time that the right-resolvent is explicitly written in full. But more importantly, finding the general right-resolvent also clearly shows the computational power of expressing in polar form the spinor field and its covariant derivative.

Avenues of research might involve applications to topology \cite{Abanov:1999qz, Golkar:2014paa, Alvarez-Gaume:1984zlq, Fabbri:2024lyu}, or for the Lie derivative \cite{GoMa, Fabbri:2023dgv}.
\vspace{10pt}

\textbf{Funding}. This work is carried out in the framework of the INFN Research Project QGSKY and funded by Next Generation EU via the project ``Geometrical and Topological effects on Quantum Matter (GeTOnQuaM)''.

\

\textbf{Data availability}. The manuscript does not have associated data in any repository.

\

\textbf{Conflict of interest}. There is no conflict of interest.
\appendix
\section{Polar Form of Spinor Fields}\label{app1}
On Euclidean surfaces, the only possible rotation is given by the exponentiation of the only generator $\boldsymbol{\sigma}_{ab}\!=\!\left[\boldsymbol{\gamma}_{a},\boldsymbol{\gamma}_{b}\right]/4$ which, for a representation of the type $\boldsymbol{\gamma}^{1}\!=\!\boldsymbol{\sigma}^{1}$, $\boldsymbol{\gamma}^{2}\!=\!\boldsymbol{\sigma}^{2}$ and $\boldsymbol{\pi}\!=\!-\boldsymbol{\sigma}^{3}$ (with $\boldsymbol{\sigma}^{1}$, $\boldsymbol{\sigma}^{2}$, $\boldsymbol{\sigma}^{3}$ Pauli matrices), is given by
\begin{eqnarray}
\boldsymbol{\sigma}^{12}\!=\!i\boldsymbol{\sigma}^{3}/2:
\end{eqnarray}
then the rotation is
\begin{eqnarray}
\boldsymbol{R}\!=\!\left(\!\begin{tabular}{cc}
$e^{i\theta/2}$ & $0$\\
$0$ & $e^{-i\theta/2}$
\end{tabular}\!\right)\label{2rotation}.
\end{eqnarray}
The most general Dirac spinor is given by
\begin{eqnarray}
\psi\!=\!\left(\!\begin{tabular}{c}
$ae^{i\alpha}$\\
$be^{i\beta}$
\end{tabular}\!\right)
\end{eqnarray}
for some module $a$ and $b$ and some phase $\alpha$ and $\beta$ real: notice that neither $a$ nor $b$ can vanish, because in such a case the spin $\overline{\psi}\boldsymbol{\gamma}^{a}\psi$ would vanish identically, and so it could not be normalized to unity. This spinor could equivalently be written according to
\begin{eqnarray}
\psi\!=\!\left(\!\begin{tabular}{c}
$ae^{i(\alpha+\beta)/2+i(\alpha-\beta)/2}$\\
$be^{i(\alpha+\beta)/2-i(\alpha-\beta)/2}$
\end{tabular}\!\right)\!=\!e^{i(\alpha+\beta)/2}\left(\!\begin{tabular}{cc}
$e^{i(\alpha-\beta)/2}$ & $0$\\
$0$ & $e^{-i(\alpha-\beta)/2}$
\end{tabular}\!\right)\!\left(\!\begin{tabular}{c}
$a$\\
$b$
\end{tabular}\!\right)
\end{eqnarray}
with the remaining spinor having two real components, which can always be written like
\begin{eqnarray}
\left(\!\begin{tabular}{c}
$a$\\
$b$
\end{tabular}\!\right)\!=\!\phi e^{-\frac{1}{2}\eta\boldsymbol{\pi}}\left(\!\begin{tabular}{c}
$1$\\
$1$
\end{tabular}\!\right)
\end{eqnarray}
for $\phi\!=\!\sqrt{ab}$ and $\eta\!=\!\ln{(a/b)}$ real, and thus $a$ and $b$ of the same sign. In this case
\begin{eqnarray}
\psi\!=\!\phi e^{-\frac{1}{2}\eta\boldsymbol{\pi}}e^{i(\alpha+\beta)/2}
\left(\!\begin{tabular}{cc}
$e^{i(\alpha-\beta)/2}$ & $0$\\
$0$ & $e^{-i(\alpha-\beta)/2}$
\end{tabular}\!\right)\!\left(\!\begin{tabular}{c}
$1$\\
$1$
\end{tabular}\!\right)
\end{eqnarray}
due to obvious commutativity properties. The matrix is just (\ref{2rotation}) for an angle $\theta\!=\!\alpha\!-\!\beta$ and when it is collected with the global unitary phase into the operator $\boldsymbol{L}^{-1}$ we can finally write
\begin{eqnarray}
\psi\!=\!\phi e^{-\frac{1}{2}\eta\boldsymbol{\pi}}\boldsymbol{L}^{-1}\left(\!\begin{tabular}{c}
$1$\\
$1$
\end{tabular}\!\right)
\end{eqnarray}
which is just (\ref{spinor2}) of section \ref{ESu}. If instead $a$ and $b$ have opposite sign, say $a$ negative and $b$ positive, we can write
\begin{eqnarray}
\left(\!\begin{tabular}{c}
$a$\\
$b$
\end{tabular}\!\right)\!=\!\boldsymbol{\pi}\left(\!\begin{tabular}{c}
$-a$\\
$b$
\end{tabular}\!\right)
\end{eqnarray}
so that now $-a$ and $b$ are both positive, and the above argument would still work. Thus, it is only up to the discrete transformation $\psi\!\rightarrow\!\boldsymbol{\pi}\psi$ that we can prove the validity of (\ref{spinor2}) in section \ref{ESu}.

On Lorentzian surfaces, one can take the representation $\boldsymbol{\gamma}^{0}\!=\!\boldsymbol{\sigma}^{3}$, $\boldsymbol{\gamma}^{1}\!=\!i\boldsymbol{\sigma}^{1}$ and $\boldsymbol{\gamma}^{2}\!=\!-i\boldsymbol{\sigma}^{2}$, giving
\begin{eqnarray}
\boldsymbol{\sigma}^{01}\!=\!-\boldsymbol{\sigma}^{2}/2\ \ \ \ \ \ \ \ \ \ \ \
\boldsymbol{\sigma}^{02}\!=\!-\boldsymbol{\sigma}^{1}/2\ \ \ \ \ \ \ \ \ \ \ \ \ \ \ \
\boldsymbol{\sigma}^{12}\!=\!i\boldsymbol{\sigma}^{3}/2:
\end{eqnarray}
then we have two boosts and one rotation given by
\begin{eqnarray}
\boldsymbol{B}_{1}\!=\!\left(\!\begin{tabular}{cc}
$\cosh{\varphi_{1}/2}$ & $-\sinh{\varphi_{1}/2}$\\
$-\sinh{\varphi_{1}/2}$ & $\cosh{\varphi_{1}/2}$
\end{tabular}\!\right)\label{3boost1}\ \ \ \ \ \ \ \
\boldsymbol{B}_{2}\!=\!\left(\!\begin{tabular}{cc}
$\cosh{\varphi_{2}/2}$ & $i\sinh{\varphi_{2}/2}$\\
$-i\sinh{\varphi_{2}/2}$ & $\cosh{\varphi_{2}/2}$
\end{tabular}\!\right)\label{3boost2}\ \ \ \ \ \ \ \ \ \ \ \
\boldsymbol{R}\!=\!\left(\!\begin{tabular}{cc}
$e^{i\theta/2}$ & $0$\\
$0$ & $e^{-i\theta/2}$
\end{tabular}\!\right)\label{3rotation}.
\end{eqnarray}
The most general Dirac spinor can be written as before, with the two phases encoded within a combination of global unitary phase and rotation, leaving a spinor with two real components: notice that these two real components cannot be equal, because in this case the velocity $\overline{\psi}\boldsymbol{\gamma}^{a}\psi$ would be space-like, and it could not be normalized to unity. Acting on this spinor with $\boldsymbol{B}_{1}$ would give
\begin{eqnarray}
\left(\!\begin{tabular}{cc}
$\cosh{\varphi_{1}/2}$ & $-\sinh{\varphi_{1}/2}$\\
$-\sinh{\varphi_{1}/2}$ & $\cosh{\varphi_{1}/2}$
\end{tabular}\!\right)\!\left(\!\begin{tabular}{c}
$a$\\
$b$
\end{tabular}\!\right)\!=\!\left(\!\begin{tabular}{cc}
$a\cosh{\varphi_{1}/2}\!-\!b\sinh{\varphi_{1}/2}$\\
$-a\sinh{\varphi_{1}/2}\!+\!b\cosh{\varphi_{1}/2}$
\end{tabular}\!\right)
\end{eqnarray}
in which, on the right-hand side, if $|a/b|\!<\!1$ there always exists a rapidity such that $\tanh{\varphi_{1}/2}\!=\!a/b$ so that the upper component vanishes, and instead if $|a/b|\!>\!1$ there always exists a rapidity such that $\tanh{\varphi_{1}/2}\!=\!b/a$ so that the lower component vanishes. Either way, one of the two components can always be set to zero, getting either
\begin{eqnarray}
&\!\psi\!=\!\phi \boldsymbol{L}^{-1}\left(\!\begin{tabular}{c}
$0$\\
$1$
\end{tabular}\!\right)\ \ \ \ \ \ \ \ \mathrm{or}\ \ \ \ \ \ \ \
\psi\!=\!\phi \boldsymbol{L}^{-1}\left(\!\begin{tabular}{c}
$1$\\
$0$
\end{tabular}\!\right)
\end{eqnarray}
with $\phi$ a real function. Notice that the sign is no longer a problem, since $a$ and $b$ of opposite sign would simply result in the rapidity having opposite sign. These are those given by (\ref{spinor3}) in section \ref{LSu}.

In Lorentzian spaces, the reasoning is analogous. Interested readers can check \cite{jl1,jl2}.
\section{Tensorial Connection and Differential Operator}\label{app2}
The general definition of covariant derivative for spinor fields is given by
\begin{eqnarray}
\boldsymbol{\nabla}_{\mu}\psi\!=\partial_{\mu}\psi
\!+\!\frac{1}{2}C_{ab\mu}\boldsymbol{\sigma}^{ab}\psi\!+\!iqA_{\mu}\psi
\label{der}
\end{eqnarray}
where $A_{\mu}$ is the gauge potential for fields  charge $q$ and $C_{ab\mu}$ is the spin-connection of the manifold (we assume that the reader is familiar with these basic concepts of gauge theory and differential geometry). This is valid in any dimension.

On Euclidean surfaces, (\ref{der}) is to be applied for the polar spinor (\ref{spinor2}), giving
\begin{gather}
\nonumber
\boldsymbol{\nabla}_{\mu}\psi\!=\partial_{\mu}\left[\phi\ e^{-\frac{1}{2}\eta\boldsymbol{\pi}}
\boldsymbol{L}^{-1}\left(\!\begin{tabular}{c}
$1$\\
$1$
\end{tabular}\!\right)\right]
\!+\!\frac{1}{2}C_{ab\mu}\boldsymbol{\sigma}^{ab}\psi\!+\!iqA_{\mu}\psi=\\
\nonumber
=\partial_{\mu}\phi\ e^{-\frac{1}{2}\eta\boldsymbol{\pi}}
\boldsymbol{L}^{-1}\left(\!\begin{tabular}{c}
$1$\\
$1$
\end{tabular}\!\right)
\!+\!\phi\ \partial_{\mu}e^{-\frac{1}{2}\eta\boldsymbol{\pi}}
\boldsymbol{L}^{-1}\left(\!\begin{tabular}{c}
$1$\\
$1$
\end{tabular}\!\right)
\!+\!\phi\ e^{-\frac{1}{2}\eta\boldsymbol{\pi}}
\partial_{\mu}\boldsymbol{L}^{-1}\left(\!\begin{tabular}{c}
$1$\\
$1$
\end{tabular}\!\right)
\!+\!\frac{1}{2}C_{ab\mu}\boldsymbol{\sigma}^{ab}\psi\!+\!iqA_{\mu}\psi=\\
=\nabla_{\mu}\ln{\phi}\ \psi
\!-\!\frac{1}{2}\nabla_{\mu}\eta\boldsymbol{\pi}\psi
\!-\!\phi\ e^{-\frac{1}{2}\eta\boldsymbol{\pi}}
(\boldsymbol{L}^{-1}\partial_{\mu}\boldsymbol{L})
\boldsymbol{L}^{-1}\left(\!\begin{tabular}{c}
$1$\\
$1$
\end{tabular}\!\right)
\!+\!\frac{1}{2}C_{ab\mu}\boldsymbol{\sigma}^{ab}\psi\!+\!iqA_{\mu}\psi
\end{gather}
since $\phi$ and $\eta$ are scalars. The object $\boldsymbol{L}^{-1}\partial_{\mu}\boldsymbol{L}$ can be explicitly computed, because we know that it is the product of a global unitary phase times a rotation of the form (\ref{2rotation}) as
\begin{eqnarray}
\boldsymbol{L}\!=\!e^{iq\alpha}\left(\!\begin{tabular}{cc}
$e^{i\theta/2}$ & $0$\\
$0$ & $e^{-i\theta/2}$
\end{tabular}\!\right).
\end{eqnarray}
Thus we find
\begin{gather}
\!\!\!\!\partial_{\mu}\boldsymbol{L}
\!=\!\partial_{\mu}e^{iq\alpha}\left(\!\begin{tabular}{cc}
$e^{i\theta/2}$ & $0$\\
$0$ & $e^{-i\theta/2}$
\end{tabular}\!\right)
\!+\!e^{iq\alpha}\partial_{\mu}\left(\!\begin{tabular}{cc}
$e^{i\theta/2}$ & $0$\\
$0$ & $e^{-i\theta/2}$
\end{tabular}\!\right)
\!=\!iq\partial_{\mu}\alpha e^{iq\alpha}\left(\!\begin{tabular}{cc}
$e^{i\theta/2}$ & $0$\\
$0$ & $e^{-i\theta/2}$
\end{tabular}\!\right)
\!+\!e^{iq\alpha}i\partial_{\mu}\theta/2\left(\!\begin{tabular}{cc}
$e^{i\theta/2}$ & $0$\\
$0$ & $-e^{-i\theta/2}$
\end{tabular}\!\right)
\end{gather}
and
\begin{eqnarray}
\boldsymbol{L}^{-1}\!=\!e^{-iq\alpha}\left(\!\begin{tabular}{cc}
$e^{-i\theta/2}$ & $0$\\
$0$ & $e^{i\theta/2}$
\end{tabular}\!\right)
\end{eqnarray}
and therefore
\begin{gather}
\boldsymbol{L}^{-1}\partial_{\mu}\boldsymbol{L}
\!=\!iq\partial_{\mu}\alpha\mathbb{I}
\!+\!i\partial_{\mu}\theta/2\left(\!\begin{tabular}{cc}
$1$ & $0$\\
$0$ & $-1$
\end{tabular}\!\right).
\end{gather}
This can be written as
\begin{gather}
\boldsymbol{L}^{-1}\partial_{\mu}\boldsymbol{L}
\!=\!iq\partial_{\mu}\alpha\mathbb{I}
\!+\!\partial_{\mu}\theta\boldsymbol{\sigma}^{12}
\!\equiv\!iq\partial_{\mu}\alpha\mathbb{I}
\!+\!\frac{1}{2}\partial_{\mu}\theta_{ab}\boldsymbol{\sigma}^{ab}
\end{gather}
were the notation $\theta\!=\!\theta_{12}\!=\!-\theta_{21}$ was introduced. We have then
\begin{gather}
\boldsymbol{\nabla}_{\mu}\psi\!=\nabla_{\mu}\ln{\phi}\ \psi
\!-\!\frac{1}{2}\nabla_{\mu}\eta\boldsymbol{\pi}\psi
\!-\!\phi\ e^{-\frac{1}{2}\eta\boldsymbol{\pi}}
(iq\partial_{\mu}\alpha\mathbb{I}
\!+\!\frac{1}{2}\partial_{\mu}\theta_{ab}\boldsymbol{\sigma}^{ab})
\boldsymbol{L}^{-1}\left(\!\begin{tabular}{c}
$1$\\
$1$
\end{tabular}\!\right)
\!+\!\frac{1}{2}C_{ab\mu}\boldsymbol{\sigma}^{ab}\psi\!+\!iqA_{\mu}\psi
\end{gather}
and because $\boldsymbol{\pi}$ and $\boldsymbol{\sigma}^{ab}$ commute
\begin{gather}
\boldsymbol{\nabla}_{\mu}\psi\!=\nabla_{\mu}\ln{\phi}\ \psi
\!-\!\frac{1}{2}\nabla_{\mu}\eta\boldsymbol{\pi}\psi
\!-\!iq(\partial_{\mu}\alpha\!-\!A_{\mu})\psi
\!-\!\frac{1}{2}(\partial_{\mu}\theta_{ab}\!-\!C_{ab\mu})\boldsymbol{\sigma}^{ab}\psi.
\end{gather}
Calling
\begin{gather}
q(\partial_{\mu}\alpha\!-\!A_{\mu})\!=\!P_{\mu}\\
\partial_{\mu}\theta_{ab}\!-\!C_{ab\mu}\!=\!R_{ab\mu}
\end{gather}
we can write
\begin{gather}
\boldsymbol{\nabla}_{\mu}\psi\!=\!(\nabla_{\mu}\ln{\phi}\mathbb{I}
\!-\!\frac{1}{2}\nabla_{\mu}\eta\boldsymbol{\pi}
\!-\!iP_{\mu}\mathbb{I}\!-\!\frac{1}{2}R_{ab\mu}\boldsymbol{\sigma}^{ab})\psi
\end{gather}
which is (\ref{decspinder2}) of section \ref{ESu}.

On Lorentzian surfaces, (\ref{der}) is to be computed for the polar spinor (\ref{spinor3}), but the result will not change, neither for the way it is derived, nor in its final structure, and one obtains (\ref{decspinder3}) of section \ref{LSu}.

In Lorentzian spaces, when (\ref{der}) is calculated for the polar spinor (\ref{spinor4}), one gets (\ref{decspinder4}) of section \ref{LSp}.

The only point that need be put in better evidence is the fact that the momentum and the tensorial connection are real tensors. This was discussed, for Lorentzian spaces, in \cite{Fabbri:2018crr}, with a proof in \cite{Fabbri:2024avj}.
\section{Covariant Decomposition of the Right-Resolvent Operator}\label{app3}
The general procedure to obtain the right-resolvent operator $\boldsymbol{R}$, and in particular its integrability conditions, very straightforward in the lower-dimensional cases, becomes elaborated in the physical space-time, where simplifications are repeatedly needed to keep expressions from extending too much. For this purpose, a trick is to work with traces.

In fact, we know that of all Clifford matrices, only the identity has non-zero trace. Then, because
\begin{gather}
\boldsymbol{\gamma}^{i}\boldsymbol{\gamma}^{j}\!=\!\eta^{ij}\mathbb{I}\!+\!2\boldsymbol{\sigma}^{ij}
\end{gather}
we have that $\mathrm{tr}(\boldsymbol{\gamma}^{i}\boldsymbol{\gamma}^{j})\!=\!4\eta^{ij}$. Similarly
\begin{gather}
\boldsymbol{\gamma}^{i}\boldsymbol{\gamma}^{j}\boldsymbol{\gamma}^{k}
\!=\!\boldsymbol{\gamma}^{i}\eta^{jk}
\!-\!\boldsymbol{\gamma}^{j}\eta^{ik}
\!+\!\boldsymbol{\gamma}^{k}\eta^{ij}
\!-\!i\varepsilon^{ijka}\boldsymbol{\gamma}^{a}\boldsymbol{\pi}
\end{gather}
and so the trace of products of three Clifford matrices is always zero. However, since
\begin{gather}
\boldsymbol{\gamma}^{a}\boldsymbol{\gamma}^{b}\boldsymbol{\gamma}^{c}\boldsymbol{\gamma}^{d}
\!=\!(\eta^{ab}\eta^{cd}\!-\!\eta^{ac}\eta^{bd}\!+\!\eta^{ad}\eta^{bc})\mathbb{I}
\!+\!2(\eta^{ab}\boldsymbol{\sigma}^{cd}
\!-\!\eta^{ac}\boldsymbol{\sigma}^{bd}
\!+\!\eta^{ad}\boldsymbol{\sigma}^{bc}
\!+\!\eta^{cd}\boldsymbol{\sigma}^{ab}
\!-\!\eta^{bd}\boldsymbol{\sigma}^{ac}
\!+\!\eta^{bc}\boldsymbol{\sigma}^{ad})
\!+\!i\varepsilon^{abcd}\boldsymbol{\pi}
\end{gather}
then $\mathrm{tr}(\boldsymbol{\gamma}^{a}\boldsymbol{\gamma}^{b}\boldsymbol{\gamma}^{c}\boldsymbol{\gamma}^{d})\!=\!4(\eta^{ab}\eta^{cd}\!-\!\eta^{ac}\eta^{bd}\!+\!\eta^{ad}\eta^{bc})$ and $\mathrm{tr}(\boldsymbol{\gamma}^{a}\boldsymbol{\gamma}^{b}\boldsymbol{\gamma}^{c}\boldsymbol{\gamma}^{d}\boldsymbol{\pi})\!=\!4i\varepsilon^{abcd}$.

Now, expression (\ref{exp}) is explicitly written as
\begin{eqnarray}
\nonumber
&i\nabla_{k}S\boldsymbol{\gamma}^{k}
\!-\!\nabla_{k}P\boldsymbol{\gamma}^{k}\boldsymbol{\pi}
\!+\!i\nabla_{k}V_{a}\boldsymbol{\gamma}^{k}\boldsymbol{\gamma}^{a}
\!+\!i\nabla_{k}A_{a}\boldsymbol{\gamma}^{k}\boldsymbol{\gamma}^{a}\boldsymbol{\pi}
\!-\!\nabla_{k}T_{ab}\boldsymbol{\gamma}^{k}\boldsymbol{\sigma}^{ab}-\\
\nonumber
&-\frac{i}{4}R_{cdk}S
\boldsymbol{\gamma}^{k}\boldsymbol{\gamma}^{c}\boldsymbol{\gamma}^{d}
\!+\!\frac{1}{4}R_{cdk}P
\boldsymbol{\gamma}^{k}\boldsymbol{\gamma}^{c}\boldsymbol{\gamma}^{d}\boldsymbol{\pi}
\!-\!\frac{i}{4}R_{cdk}V_{a}
\boldsymbol{\gamma}^{k}\boldsymbol{\gamma}^{a}\boldsymbol{\gamma}^{c}\boldsymbol{\gamma}^{d}
\!-\!\frac{i}{4}R_{cdk}A_{a}
\boldsymbol{\gamma}^{k}\boldsymbol{\gamma}^{a}\boldsymbol{\gamma}^{c}\boldsymbol{\gamma}^{d}\boldsymbol{\pi}
\!+\!\frac{1}{4}R_{cdk}T_{ab}
\boldsymbol{\gamma}^{k}\boldsymbol{\sigma}^{ab}\boldsymbol{\gamma}^{c}\boldsymbol{\gamma}^{d}+\\
\nonumber
&+P_{k}S\boldsymbol{\gamma}^{k}
\!+\!iP_{k}P\boldsymbol{\gamma}^{k}\boldsymbol{\pi}
\!+\!P_{k}V_{a}\boldsymbol{\gamma}^{k}\boldsymbol{\gamma}^{a}
\!+\!P_{k}A_{a}\boldsymbol{\gamma}^{k}\boldsymbol{\gamma}^{a}\boldsymbol{\pi}
\!+\!iP_{k}T_{ab}\boldsymbol{\gamma}^{k}\boldsymbol{\sigma}^{ab}-\\
&-mS\mathbb{I}
\!-\!imP\boldsymbol{\pi}
\!-\!mV_{a}\boldsymbol{\gamma}^{a}
\!-\!mA_{a}\boldsymbol{\gamma}^{a}\boldsymbol{\pi}
\!-\!imT_{ab}\boldsymbol{\sigma}^{ab}
\!-\!\phi\cos{(\beta/2)}\mathbb{I}
\!+\!i\phi\sin{(\beta/2)}\boldsymbol{\pi}\!=\!0
\end{eqnarray}
and taking its trace one can see that only the terms
\begin{eqnarray}
&\mathrm{tr}[i\nabla_{k}V_{a}\boldsymbol{\gamma}^{k}\boldsymbol{\gamma}^{a}
\!-\!\frac{i}{4}R_{cdk}V_{a}
\boldsymbol{\gamma}^{k}\boldsymbol{\gamma}^{a}\boldsymbol{\gamma}^{c}\boldsymbol{\gamma}^{d}
\!-\!\frac{i}{4}R_{cdk}A_{a}
\boldsymbol{\gamma}^{k}\boldsymbol{\gamma}^{a}\boldsymbol{\gamma}^{c}\boldsymbol{\gamma}^{d}\boldsymbol{\pi}
\!+\!P_{k}V_{a}\boldsymbol{\gamma}^{k}\boldsymbol{\gamma}^{a}
\!-\!mS\mathbb{I}\!-\!\phi\cos{(\beta/2)}\mathbb{I}]\!=\!0
\end{eqnarray}
may contribute to the final integrability condition. Taking the traces leads to
\begin{eqnarray}
&i\nabla_{i}V^{i}\!+\!P_{i}V^{i}\!-\!\frac{i}{2}R_{i}V^{i}
\!-\!\frac{1}{4}\varepsilon^{kabc}R_{abc}A_{k}\!-\!mS\!-\!\phi\cos{(\beta/2)}\!=\!0
\end{eqnarray}
which is the first of the integrability conditions. If (\ref{exp}) is multiplied by $\boldsymbol{\pi}$, and the trace is taken, we get that only
\begin{eqnarray}
&\mathrm{tr}[i\nabla_{k}A_{a}\boldsymbol{\gamma}^{k}\boldsymbol{\gamma}^{a}
\!-\!\frac{i}{4}R_{cdk}V_{a}
\boldsymbol{\gamma}^{k}\boldsymbol{\gamma}^{a}\boldsymbol{\gamma}^{c}\boldsymbol{\gamma}^{d}\boldsymbol{\pi}
\!-\!\frac{i}{4}R_{cdk}A_{a}
\boldsymbol{\gamma}^{k}\boldsymbol{\gamma}^{a}\boldsymbol{\gamma}^{c}\boldsymbol{\gamma}^{d}
\!+\!P_{k}A_{a}\boldsymbol{\gamma}^{k}\boldsymbol{\gamma}^{a}
\!-\!imP\mathbb{I}\!+\!i\phi\sin{(\beta/2)}\mathbb{I}]\!=\!0
\end{eqnarray}
contribute. Tracing, gives
\begin{eqnarray}
&i\nabla_{k}A^{k}\!+\!P_{k}A^{k}\!-\!\frac{i}{2}R_{k}A^{k}
\!-\!\frac{1}{4}R_{ijk}V_{a}\varepsilon^{aijk}\!-\!imP\!+\!i\phi\sin{(\beta/2)}\!=\!0
\end{eqnarray}
which is the second integrability condition. Multiplying instead by $\boldsymbol{\gamma}^{i}$, $\boldsymbol{\gamma}^{i}\boldsymbol{\pi}$, $\boldsymbol{\sigma}^{ab}$, and tracing, would give the others.

\end{document}